\documentclass[a4paper,fleqn,usenatbib,useAMS]{mnras}

\usepackage{graphicx}	
\usepackage{amsmath}	
\usepackage{amssymb}	
\usepackage{multicol}        
\usepackage{bm}		
\usepackage{pdflscape}	
\usepackage{natbib}
\usepackage{times}
\usepackage{calc}
\usepackage{rotating}
\usepackage{lscape}
\bibpunct{(}{)}{;}{a}{}{,} 
\usepackage{longtable}
\usepackage{enumitem} 
\usepackage{hyperref}
\usepackage{pifont}
\usepackage{tikz}
\usepackage{xcolor}
\usepackage[title]{appendix}

\def\oversim#1#2{\lower0.5pt\vbox{\baselineskip0pt \lineskip-0.5pt
     \ialign{$\mathsurround0pt #1\hfil##\hfil$\crcr#2\crcr\sim\crcr}}}

\usepackage[T1]{fontenc}
\usepackage{ae,aecompl}

\usepackage{mathptmx}

\def\aap {{A\&A}}

\def\apjl {{ApJ}}

\def\apjs {{ApJs}}

\title[ALMA analysis of K3-35]{An attempt to determine the magnetic field configuration in the planetary nebula K\,3-35 with ALMA}

\author[L. Sabin,et al.] 
{L. Sabin$^{1}$\thanks{E-mail:lsabin@astro.unam.mx (LS)} , Q. Zhang$^{2}$, L.F. Miranda$^{3}$ and M.~A. Gómez-Muñoz$^{4,5}$\\
\\
$^{1}$Universidad Nacional Aut\'onoma de M\'exico. Instituto de Astronom\'ia. A.P. 106, 22800. Ensenada, B.C., M\'exico \\
$^{2}$Center for Astrophysics | Harvard \& Smithsonian, 60 Garden Street, Cambridge, MA 02138, USA \\
$^{3}$Instituto de Astrof\'{i}sica de Andaluc\'{i}a (IAA) -CSIC, Glorieta de la Astronom\'{i}a S/N, 18008 Granada, Spain\\
$^{4}$Departament de F{\'i}sica Qu{\`a}ntica i Astrof{\'i}sica (FQA), Universitat de Barcelona (UB),  c. Mart{\'i} i Franqu{\'e}s, 1, 08028 Barcelona, Spain\\
$^{5}$Institut de Ci{\`e}ncies del Cosmos (ICCUB), Universitat de Barcelona (UB), c. Mart{\'i} i Franqu{\'e}s, 1, 08028 Barcelona, Spain
}

\date{Last updated 2025 xx xx; in original form 2025 xx xx}
\pubyear{2025}

\begin{document}

\label{firstpage}
\pagerange{\pageref{firstpage}--\pageref{lastpage}}
\maketitle


\begin{abstract}

We examined dust polarisation within the planetary nebula (PN) K\,3-35 using the Atacama Large Millimeter/Submillimeter Array (ALMA). This investigation aimed to identify and trace the magnetic field within the PN, as it potentially plays a crucial role in shaping this bipolar nebula. Our findings include a marginal detection of the polarised region and low fractional polarisation (peaking at 1.4\%). Assuming a certain level of validity, we observed well-organised dust grains aligned along the equatorial plane of the PN, indicating a magnetic field alignment with the outflows. The limited detection of polarisation at submillimeter wavelengths in this PN and others may be attributed to a pronounced optical depth. However, our analysis of K\,3-35 with the code {\sc DUSTY} does not seem to support this idea. We also modelled the SED of K3-35, and our best-fit models included a mixture of silicates and amorphous carbon. The grains of amorphous carbon are less susceptible to alignment with the magnetic field, which could, at least partially, explain the observed low polarisation.  The models presented in this article should be considered preliminary, and a more advanced approach is needed for a more complete interpretation of the results.

\end{abstract}

\begin{keywords}
magnetic fields --- polarisation --- stars: AGB and post-AGB --- ISM: jets and outflows
planetary nebula: individual: K\,3-35
  
\end{keywords}

\section{Introduction}

Magnetic fields are known to be a key element in the formation and development of stars at various stages of their evolution, and evolved low- and intermediate-mass stars, namely Planetary Nebulae (PNe), do not seem to be an exception. In this case, magnetic fields are thought to play a role in the shaping mechanism that transforms the Asymptotic Giant Branch (AGB) spherically symmetric envelopes into axis-symmetric/nonspherical PNe. Various theoretical (see \citealt{Huarte2012},  \citealt{Blackman2001}, \citealt[and references therein]{Garcia2022}) and observational studies (see \citealt{Suarez2015}, \citealt{Duthu2017}, \citealt{Sabin2020}[and references therein], \citealt{Vlemmings2023}) have been conducted, thereby probing the importance of those fields. Still, from an observational and statistical point of view, the number of well studied ''magnetic PNe'' is still scarce, thereby hampering the precise analysis of the implications or effects of the presence of magnetic fields in these evolved stars. \\

K\,3-35 is a PN in which the presence of a nebular magnetic field has been 
reported. Some features of the OH maser emission at 1665 and, marginally, 1667 MHz present circular polarisation as indicated by \citealt{Miranda2001}(hereafter M+01). \citet{Gomez2009} identified a possible Zeeman pair in the 1665 MHz transition, which, if confirmed, would indicate a magnetic field strength of $\sim$0.9\,mG at 150\,AU from the 
center at a distance to K3-35 of 3.9\,kpc \citep{Tafoya2011}. However, the existence of Zeeman splitting in the 1665 MHz OH line (or any other of the OH transitions actually) was not confirmed by \citet{Gomezb2016}, although they mentioned a high noise level in their spectra. 
\citet{Hou2020} observed the OH excited state at 6.035\,GHz 
(see also Desmurs et al. 2010) and found an S-shaped profile in the Stokes V 
spectrum, which, if due to Zeeman-splitting, would indicate magnetic 
field strengths of $\sim$2.9 and 4.5\,mG. 

In addition to the results mentioned above, K\,3-35 also exhibits more interesting properties. 
It is a very young PN, as indicated by the presence of water maser emission in the 
nebula (M+01). The nebula is bipolar with an angular size of $\sim$6$''$, and contains a precessing 
bipolar jet that can be traced to the innermost regions of the object (M+01).
The water masers arise in a dense equatorial ring of $\sim$65\,AU in radius and at the tips 
of the bipolar jet at $\sim$3800\,AU from the center (M+01, \citealt{Uscanga2008,Tafoya2011}). The innermost regions of the bipolar jets are almost perpendicular to the water maser ring, suggesting that the ring could be involved in the collimation 
of the bipolar jet \citep{Uscanga2008}. Moreover, analysis of near- and mid-IR, and 
radio-continuum images indicates that dust is present in the core of the object and in the bipolar 
jets \citep{Blanco2014}. In particular, the image at $\lambda$\,=11.85 $\mu$m (SiC filter) 
obtained at 0.3\,arcsec resolution very likely shows the dust distribution in K\,3-35 
(\citealt{Blanco2014}, their Fig.1-Right).\\
K\,3-35 is therefore an interesting source for which the local intensity of the magnetic field has been measured. However, its configuration remains relatively unknown. The aforementioned results make K\,3-35 an ideal PN for polarisation studies using ALMA. Thus, the analysis of the dust polarisation would indicate whether the field 
is organised and follows a toroidal or polar distribution. This could, in turn, be related to the type of magnetic launching operating in the source \citep{Ceccobello2021} which would be responsible for outflow ejections.\\
In this paper, we present ALMA continuum polarisation data of K\,3-35, which have been obtained to analyse the morphology of the magnetic field in this young PN. The paper is organised as follows. In Section \S2, we present the ALMA 
polarisation data. In Section \S3, the results for the continuum emission and dust polarisation are shown. 
Finally, the discussion and conclusions are presented in Sections \S4 and \S5. \\

\begin{figure*}
\includegraphics[width=0.64\linewidth]{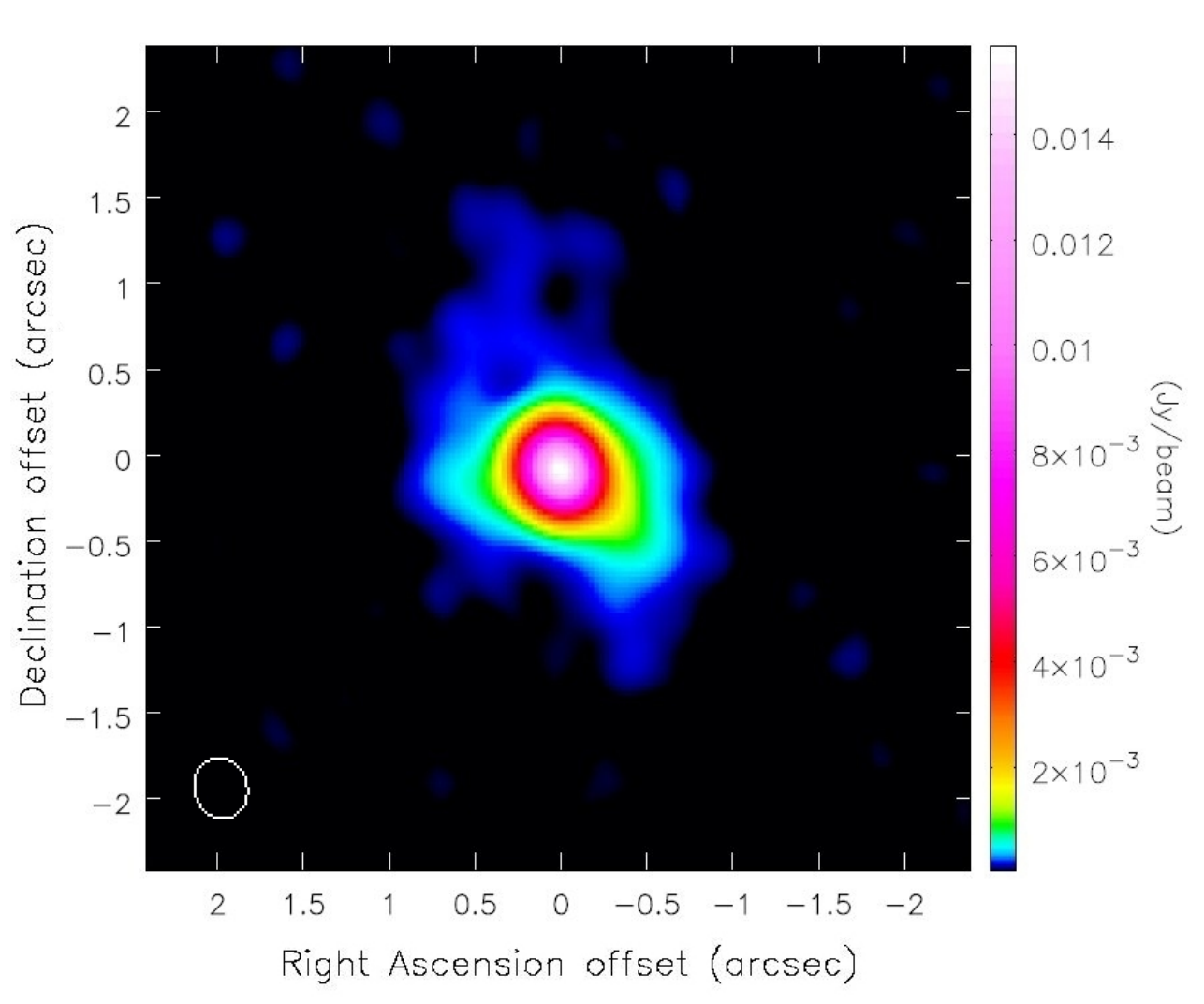}
\includegraphics[width=0.6\linewidth]{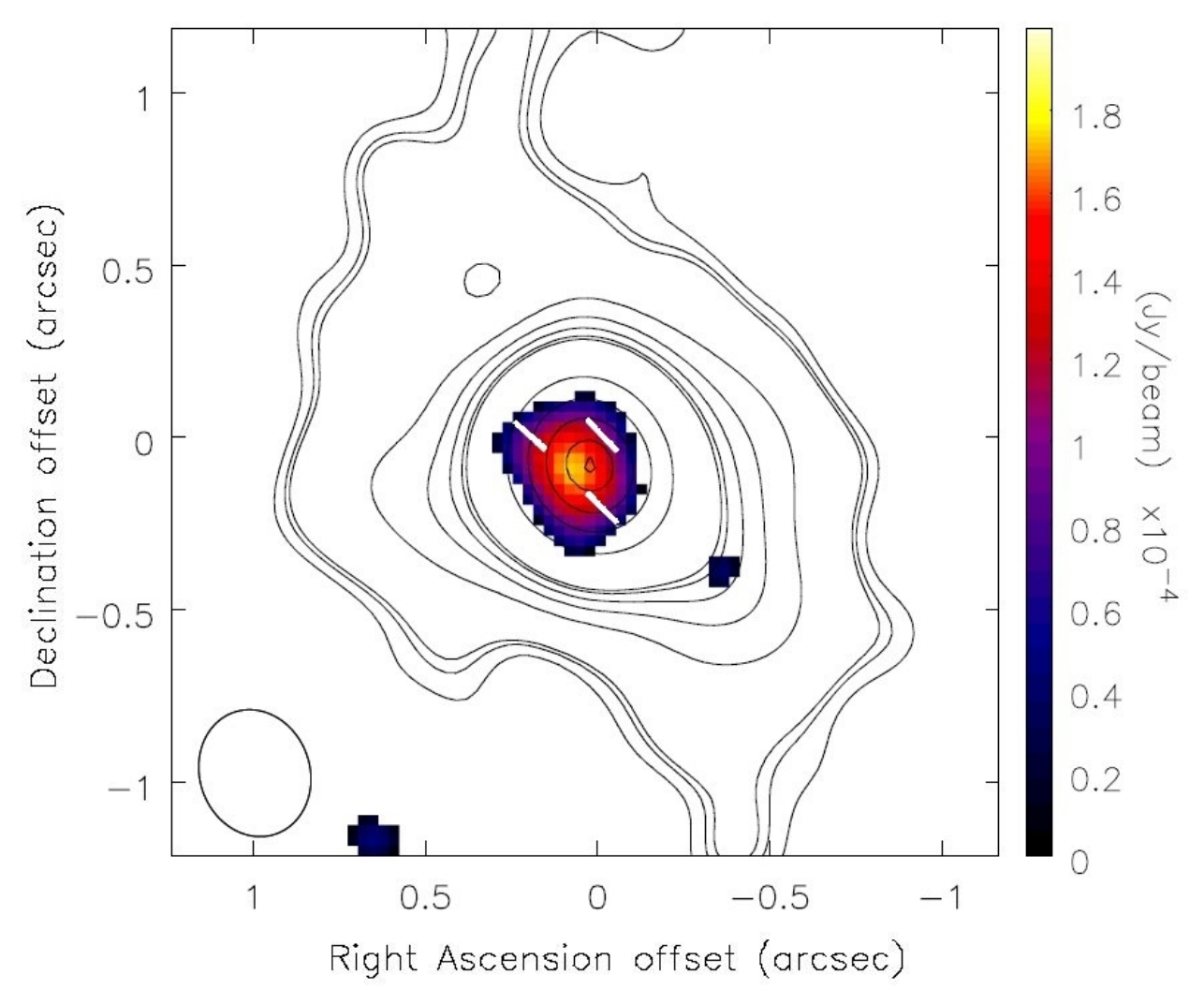}
\caption{Top: Stokes I ALMA map of K3-35 with the superimposed polarisation electric. Bottom: Polarised emission of K3-35 is shown in the colormap with superimposed magnetic field segments in white. The magnetic field orientation was determined by rotating the electric vectors by 90$^{\circ}$. The total dust continuum emission is shown as a contour plot. The relative contour levels are drawn at the levels [0.007, 0.009, 0.01, 0.03, 0.05, 0.07, 0.09, 0.1, 0.3, 0.5, 0.7, 0.9, 0.99] where 1 corresponds to 15.72 mJy/beam. The synthesized beam, with a size of 0.37\arcsec$\times$0.32\arcsec, is presented in the lower left corner}.
\label{ALMA_IB}
\end{figure*}


\section[]{Observations and data reduction}\label{obs}

The ALMA data presented in this article were acquired on 2018-09-05 using 46$\times$12m antennas (Project: 2016.1.00944.S, P.I.\,Sabin). Two runs, of 92.33 min and 74.98 min total duration, were conducted in Band 7 with a representative frequency of 348.50 GHz. The observations were performed under the C40-4 configuration with a maximum baseline of 783.5 m and a maximum recoverable scale (MRS) of 4\arcsec.
The bandpass, flux, atmosphere, and pointing calibrations were performed using the quasar J1751+0939. In addition, the atmosphere calibration included J1924-2914 and the main target K3-35. The pointing calibration also included J1924-2914 and J1925+2106. The phase calibration was performed using J1935+2031. Finally, the source J1924-2914 was also employed for polarisation calibration.\\
Data reduction was performed with the ALMA pipeline using the Common Astronomy Software Applications \citep[CASA v 5.4.0]{McMullin2007}. Both runs were calibrated separately and then combined. Imaging and polarisation calibration were performed on the combined dataset. The signal-to-noise ratio of the ALMA data was improved after three iterative rounds of phase-only self-calibration, and we used Briggs weighting with a robust factor of 0.5. As a result, we obtained images with a synthesised beam of 0.37\arcsec$\times$0.32\arcsec and a position angle PA=18.92$\degr$. The estimated uncertainty of instrumental polarisation is approximately 0.1\%.\\

\section[]{Results}\label{cont}

\subsection[]{Continuum emission}\label{cont}

The primary beam-corrected dust continuum image indicates that the submillimeter emission in K3-35 extends over $\sim$2.0\arcsec$\times$3.0\arcsec (Fig.\ref{ALMA_IB}). It displays a bright elongated central structure (east-west) with two nearly perpendicular and opposite asymmetric extensions (towards the northeast and the southwest) forming an S-shaped morphology, consistent with previous studies. The 0.86 mm map shows a peak intensity of 0.0157 Jy/beam at the coordinates $\alpha$(J2000) =19$^{h}$27$^{m}$44$^{s}$.020, $\delta(J2000) = $+21$\degr$30$\arcmin$03$\arcsec$.309 corresponding to the bright central area of the nebula, and a mean intensity of the total dust emission of $9.7 \times 10^{-4}$ Jy/beam.


\subsection[]{Dust polarisation analysis}\label{cont}

\begin{figure}
\includegraphics[width=\linewidth]{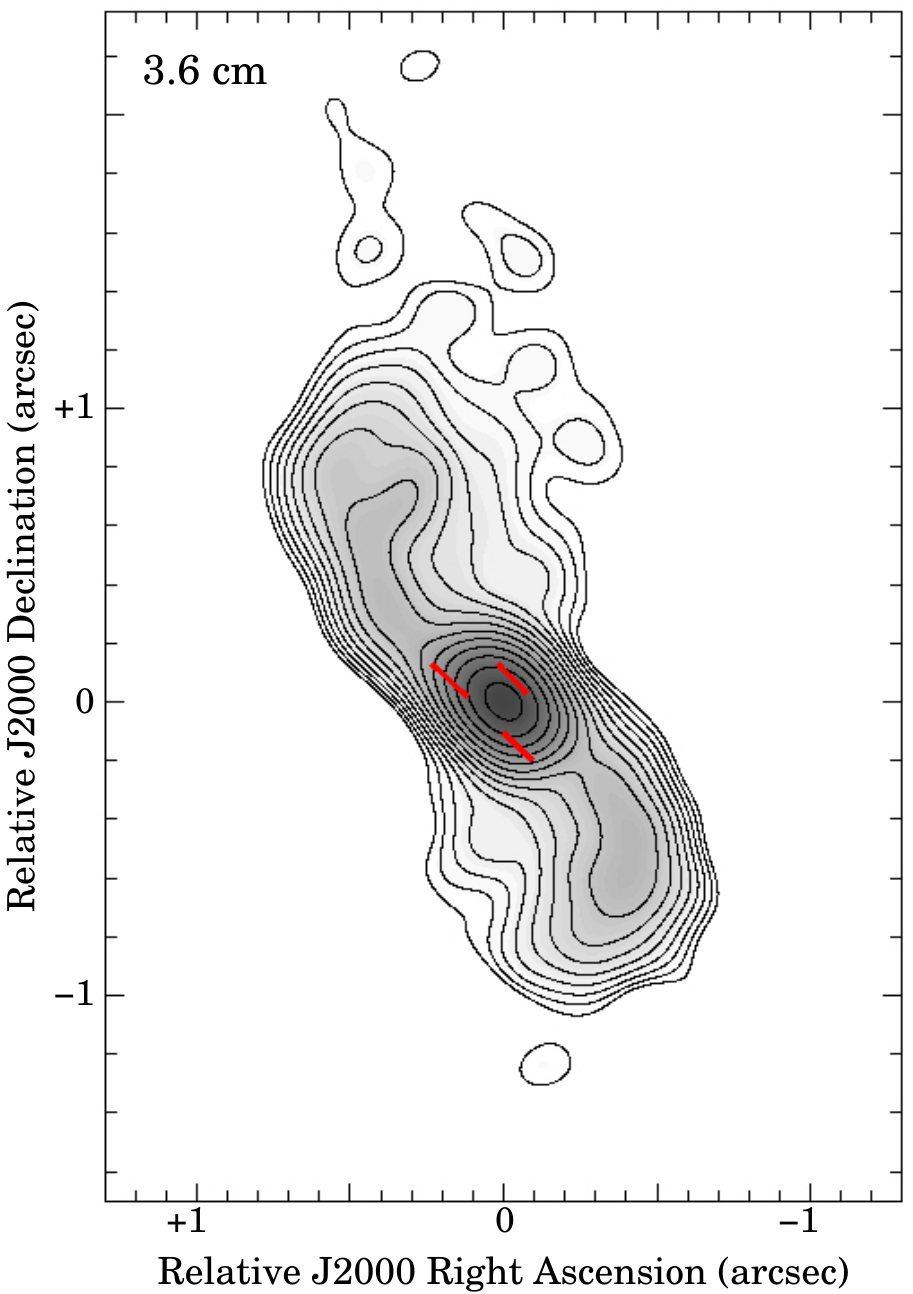}
\caption{Map of the continuum intensity at 3.6 cm of K3-35 obtained  with the VLA (adapted from M+01) superimposed with the possible distribution of the magnetic field segments from  this work.}
\label{Centimeter}
\end{figure}

In the first instance, the ALMA polarisation map was produced with a 3$\sigma$ cut (rms = 19.7 $\mu$Jy/beam) which is a typical threshold to ensure a robust detection. The peak polarisation of 1.615$\times$10$^{-4}$ Jy/beam, is located at coordinates $\alpha$(J2000) =19$^{h}$27$^{m}$44$^{s}$.024, $\delta(J2000) = $+21$\degr$30$\arcmin$03$\arcsec$.310 and does not exactly coincide with the peak of thermal emission, that is, with a very slight offset\footnote{The offset of 0.004\arcsec East is however not significant as it is well within the synthesised beam}. The mean emission over the entire $\sim$0.36$\arcsec$ $\times$0.42$\arcsec$ polarised area is $9.24 \times 10^{-5}$ Jy/beam.  The measurement of the fractional polarisation map indicates a maximum polarisation level of 1.4\% ( with a mean value of $\sim$0.9\%).\\
The size of the beam makes it clear that the polarised emission is barely resolved. Indeed, the emission area is only $\sim$1.1 times larger than that of the beam. Therefore, while polarisation is detected, it is not possible to draw any firm conclusions about the distribution of the magnetic field using dust alignment theory \citep{Lazarian2007}. If we reduce the threshold limit to 2.5$\sigma$ (see Fig.\ref{ALMA_IB}), the extent of the polarised emission is now $\sim$1.4 times that of the beam. Therefore, we will work with these data sets to investigate the approximate position of the electric ($\bar{E}$) and magnetic ($\bar{B}$) segments as well as the polarisation fraction within the nebula. \\
The electric vectors might seem organised at first sight with a mean position angle distribution of $\sim$ -44.3$^{\circ}$ (with a standard deviation of 4.1$^{\circ}$) which corresponds to the equatorial plane of K3-35. Because $\bar{B}$ is perpendicular to $\bar{E}$, our data then indicate that the magnetic field is oriented in the direction of the bipolar outflows.  This is illustrated in Fig.\ref{Centimeter} where we show the 3.6 cm radio continuum map of K3-35 with the magnetic field orientation superimposed. Magnetic fields oriented alongside the outflows have already been found (such as the case of OH 231.8+4.2, see \citealt{Sabin2014,Sabin2020}) whether being dragged or tracing a more dynamic pattern, i.e, a launching mechanism. Thus, we might be observing a case of magnetic collimation in K3-35 as well. However, the observed uniformity in the polarisation maps is more likely due to the emission being barely resolved. Additionally, since the detection of polarised emission is based on a low 2.5-sigma threshold, these results should be interpreted with caution.


\begin{table}
    \caption{Optical and IR photometric fluxes of K\,3-35.}
    \label{tab:photometry}
    \centering
    \begin{tabular}{rrc}
    \hline
    Wavelength & Flux & Band \\
    ({\micron}) & (log Jy) & \\
    \hline
0.47 & $-$4.08 & SDSS  g \\ 
0.48 & $-$4.14 & PS1  g \\ 
0.50 & $-$3.72 & GAIA3  Gbp \\ 
0.55 & $-$3.77 & Johnson  V \\ 
0.58 & $-$3.57 & ACSWFC  F606W \\ 
0.58 & $-$3.58 & GAIA3  G \\ 
0.67 & $-$3.35 & Johnson  R \\ 
0.75 & $-$3.14 & SDSS  i \\ 
0.75 & $-$3.18 & PS1  i \\ 
0.76 & $-$3.05 & GAIA3  Grp \\ 
0.77 & $-$3.13 & IPHAS  gI \\ 
0.80 & $-$3.06 & ACSWFC  F814W \\ 
0.86 & $-$3.01 & Johnson  I \\ 
0.87 & $-$2.94 & PS1  z \\ 
0.89 & $-$2.89 & SDSS  z \\ 
0.96 & $-$2.75 & PS1  y \\ 
1.24 & $-$2.18 & 2MASS  J \\ 
1.66 & $-$1.97 & 2MASS  H \\ 
2.16 & $-$1.63 & 2MASS  Ks \\ 
3.35 & $-$1.80 & WISE  W1 \\ 
4.60 & $-$1.34 & WISE  W2 \\ 
7.95 & $-$0.26 & MSX  A \\ 
8.23 & $-$0.43 & IRC  S9W \\ 
11.56 & $-$0.15 & WISE  W3 \\ 
12.07 & 0.22 & MSX  C \\ 
14.59 & 0.61 & MSX  D \\ 
17.61 & 0.78 & IRC  L18W \\ 
21.02 & 1.16 & MSX  E \\ 
21.74 & 1.47 & IRAS  25{\micron} \\ 
22.09 & 0.97 & WISE  W4 \\ 
52.14 & 1.68 & IRAS  60{\micron} \\ 
62.95 & 1.58 & FIS  N60 \\ 
76.89 & 1.29 & FIS  WIDE-S \\ 
95.48 & 1.26 & IRAS  100{\micron} \\ 
850.00 & $-$1.629 & ALMA-345GHz \\ 
\hline
    \end{tabular}
\end{table}

\section[]{Discussion}\label{disc}


\subsection[]{The lack of polarisation in K3-35}

The marginal detection of the polarised emission in K\,3-35 resembles that of CRL 618 \citep{Sabin2014} and to a lesser extent that of Frosty Leo \citep{Sabin2019}, both observed with the Submillimeter Array (SMA) around 345 GHz with a similar synthesised beam of $\sim$2.2\arcsec $\times$1.9\arcsec. 
 Assuming that the observed pattern accurately represents the true configuration of the magnetic field, similar to the case of CRL 618, the dust grains in K 3-35 would be organised and aligned along the equatorial plane of the planetary nebula. This alignment would be consistent with the presence of a magnetic field oriented along the outflows. In addition, it should also be pointed out that this interpretation assumes that dust polarisation is principally linked to the presence of magnetic fields and not to self-scattering \citep{Kataoka2015}.

Our primary aim was to identify and map the magnetic field within the inner region of the nebula. This was undertaken to gain a better understanding of the magnetic launching phenomenon. However, as with other evolved stars previously mentioned, such an analysis is not clear in the case of K\,3-35, because of the absence of significant polarised emission. We identified three factors that could explain this pattern. First, the inner areas typically contain more dust particles (in some cases a ``dark lane'' can be observed), resulting in increased  optical depth. The generally low polarisation at sub-millimeter wavelengths may imply that the source is relatively optically dense or thick, impeding our ability to obtain information on polarisation levels.
Then, the lack of polarisation at submillimeter wavelengths could also be linked to the size of the dust particles in the source. Because the degree of polarisation is contingent on the alignment of dust grains (or, stated differently, a low grain alignment level will result in a low fractional polarisation), and assuming the RAdiative Torques (RAT) alignment theory, smaller grains are less likely to align efficiently. Finally, the intricate physical conditions within the source, such as depolarisation processes due to collisions for example 
(which ultimately affects the grain size and the alignment) could impede polarisation detection.\\

\begin{figure*}
    \centering
    \includegraphics[width=0.48\textwidth]{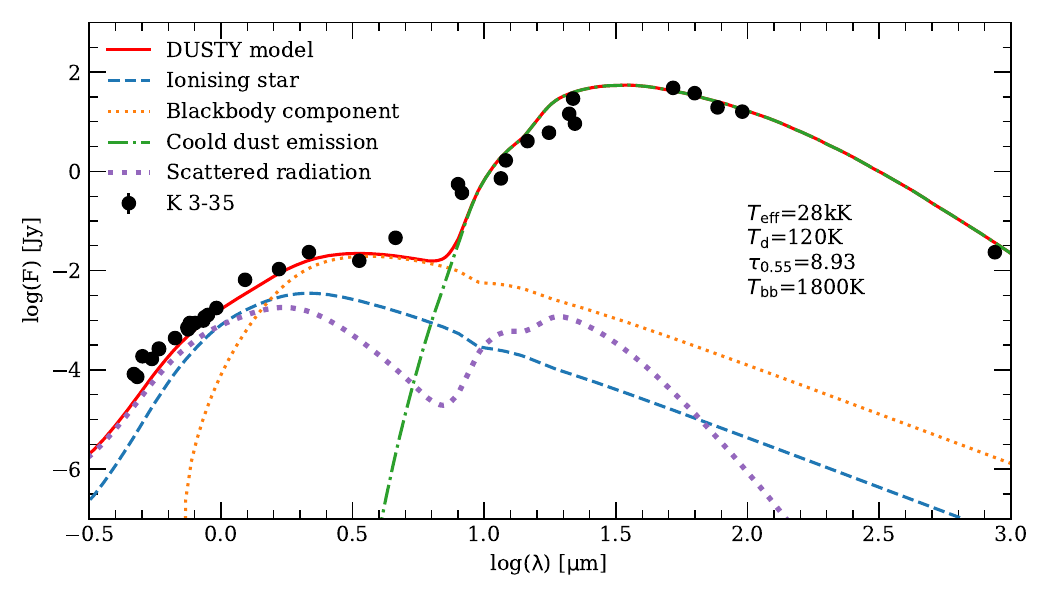}
    \includegraphics[width=0.48\textwidth]{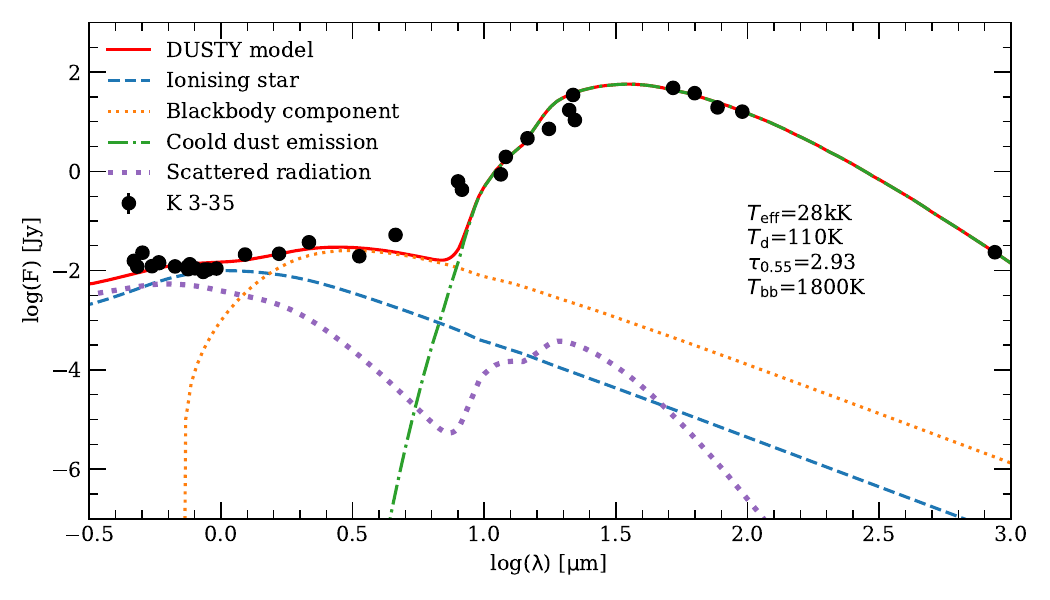}
    \includegraphics[width=0.48\textwidth]{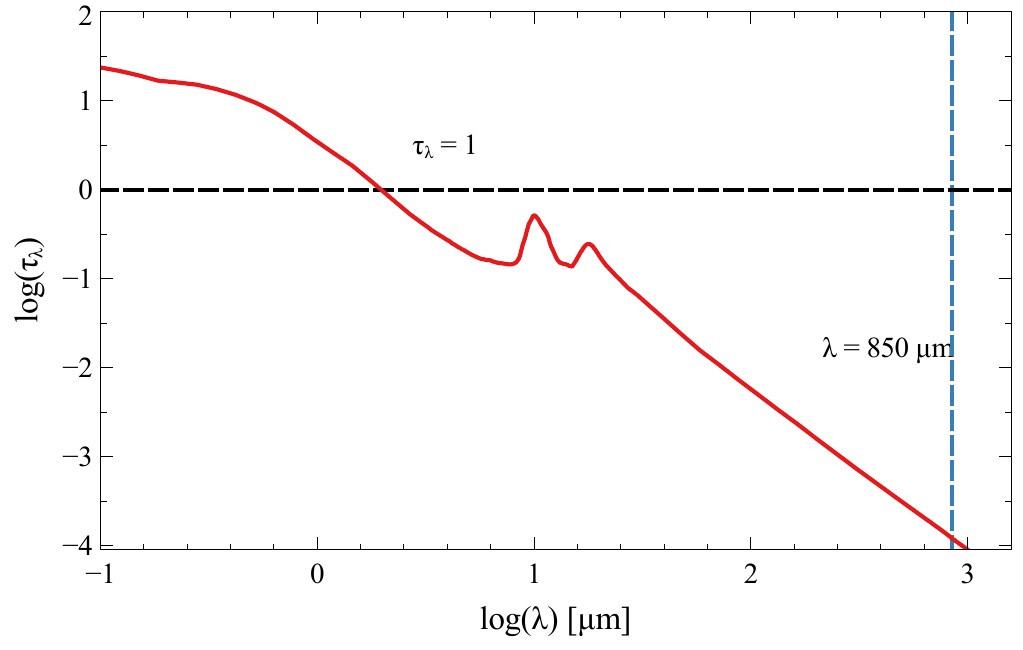}
    \includegraphics[width=0.48\textwidth]{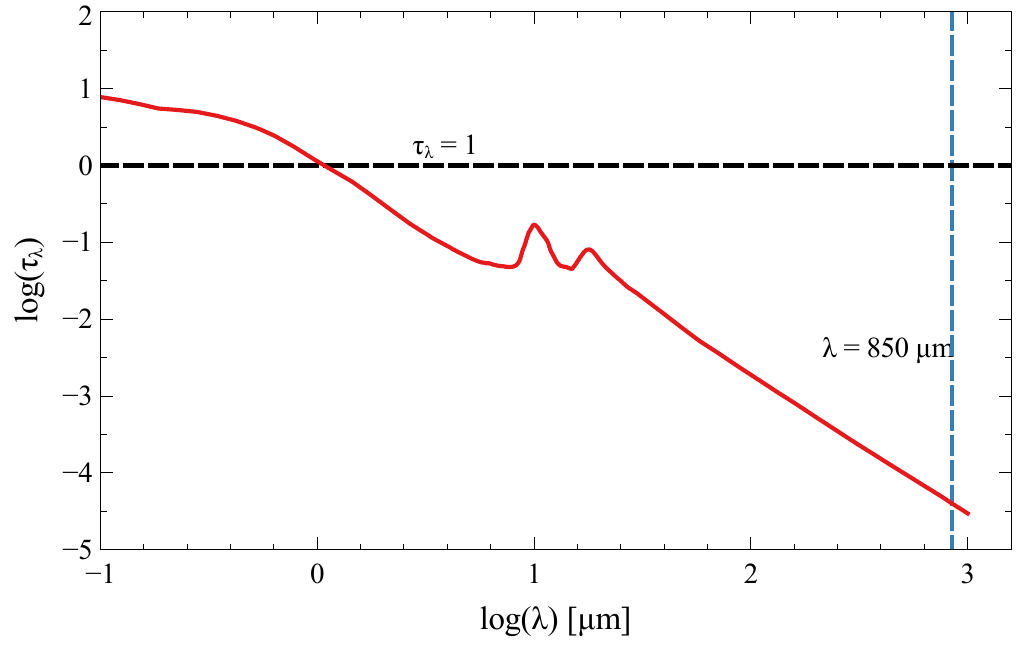}
    \caption{Top: {\sc DUSTY}  model, with its different components, based on the reddened (left) and unreddened (right) photometry of K3-35. Bottom: Corresponding optical depth distributions with values of 1.21e-4 (left) and 3.97e-5 (right) at 850$\mu$m.}
    \label{fig:dusty_model}
\end{figure*}


\subsection{{\sc DUSTY} modelling}

In order to examine if the optical depth is indeed the cause of low polarisation in K\,3-35, we computed a radiative transfer model using the 1-D radiative transfer code {\sc DUSTY} \citep[v4;][]{dusty} which allows us to determine the optical depth at a fiducial wavelength, $\tau_\mathrm{V}$ (at 0.55\,{\micron}), by using the dust temperature $T_\mathrm{d}$, at the inner shell boundary $R_\mathrm{in}$, and the effective temperature of the star $T_\mathrm{eff}$ as inputs. The code is based on scaling relations appropriate for symmetric spherical dust shells in which the thickness; $Y=R_\mathrm{out}/R_\mathrm{in}$, and the density profiles are also defined as inputs.

To obtain the optical depth as a function of wavelength for the PN K\,3-35, we first downloaded the best photometric data available in the different optical and IR databases (see Table\,\ref{tab:photometry}; note that the mean values were taken for multiple observations). Then, a reduced $\chi^{2}$ minimisation was applied to {\sc DUSTY} models to match the observed data. {\sc DUSTY} was invoked by considering a steady-state wind with a typical density power law distribution as $r^{-2}$. A chemical dust composition of silicates \citep{Draine1984} was considered, as it was appropriate for O-rich objects and also due to the detection of SiC dust distribution feature by \citet{Blanco2014},  with a canonical ISM grain size distribution ($a_\mathrm{min}$=0.005\,{\micron}, $a_\mathrm{max}$=0.25\,{\micron}, and power law of $-3.5$). As K3-35 is known to be a young PN or post-AGB, we computed grids of $T_\mathrm{eff}$ from 2000 to 35000\,K in steps of 1000\,K assuming a blackbody model with an inner dust temperature ($T_\mathrm{d}$) between 100 and 900\,K in steps of 100\,K. We also considered optical depths, $\tau_\mathrm{V}$, from 0 to 10 in logarithmic steps. Once approaching the best fitting model, a finer grid was created. For each model, we convolved the IR photometric transmission bands ($>$1\,{\micron}) and compared them to the observed flux. For wavelengths $<$1\,{\micron} we interpolated the theoretical {\sc DUSTY} spectra to the observed magnitudes at each wavelength. The results of the  {\sc DUSTY} modelling are shown in Fig.\ref{fig:dusty_model}) and we note that a black body with $T_\mathrm{bb}=1800$\,K is additionally required to reproduce the SED (see below).

According to \citet{Miranda2000}, K\,3-35 is located towards the molecular cloud L\,755, making it difficult to separate the ISM from circumstellar extinction (IR extinction is poorly understood in these conditions). In this case, we fitted both the reddened and unreddened observed SED (see the left- and right-hand panels of Figure\,\ref{fig:dusty_model}, respectively) where $E(B-V)=1.52\pm0.3$\,mag was obtained by \citealt{Vickers2015}. It is to be noted that our best-fit models include a mixture of silicates
and amorphous carbon \citep{Rouleau1991} for an H$_2$O/OH maser emission object. Although the presence of a mixture of silicates and amorphous carbon of such an apparently O-rich post-AGB star is unexpected, such double chemistry has been observed in other maser emitting PNe \citep{Miranda2021,Cala2024}. Thus, K\,3-35 has likely experienced or may currently be experiencing a conversion to a C-rich object \citep{Bunzuel2009}. Furthermore, the existence of amorphous carbon (according to the best-fit models) may contribute to the low polarisation, because these grains are less likely to align with the magnetic field. As mentioned above, our best models notably include a black body component with $T_\mathrm{bb}=1800$\,K (Fig.\ref{fig:dusty_model}), which suggests the possible presence of either a (very) cool binary companion or hot dust.\\

The analysis of optical depth derived from our models (see Fig.\ref{fig:dusty_model}
, bottom row) indicates that at 850 $\mu$m, the medium is optically thin ($\tau \ll 1$). Thus, the low detected polarisation level is unlikely to be due to the high optical depth of the source. However, we found no information in the literature regarding the implications of such low optical depth (down to $10^{-5}$) in evolved stars.\\
The models used to represent the emission in K\,3-35 generally employed small dust grains, whose alignment efficiency may not be optimal for detecting polarisation.\\
However, it is important to emphasize that our models, based on a 1D spherical symmetry approach, are limited in fully reproducing the observed properties of the nebula. Therefore, this modelling should be considered a preliminary attempt, and more advanced approaches (ideally incorporating turbulence, for example)  will be necessary to analyse the depolarisation effect due to the conditions at the core of K\,3-35 and, more generally, for a more complete interpretation of the results. Such analysis is beyond the scope of the present study.

\section{Conclusions}

We investigated the polarised emission in the young PN K3-35 with ALMA, aiming to detect the distribution of magnetic fields. The polarised emission is marginally detected (with a peak polarisation fraction of 1.4\%) and is barely resolved in the object, resulting in an imprecise estimate of the polarisation vector distribution. Nevertheless, one possibility is that the electric vectors are indeed aligned along the equatorial plane; therefore, the magnetic field might be concordant with the bipolar outflows. We also explored the possible reasons for the low polarisation levels using the code {\sc DUSTY} and found that the  optical depth was unlikely to be the cause (at the observed wavelength), and that the dust size might play a role in the observed low polarisation. On the other hand, our SED model includes a mixture of silicates and amorphous carbon; the latter would probably align less with the magnetic field, contributing to the observed low polarisation.

\section*{Acknowledgements}

We thank the anonymous referee for providing insightful comments which strengthened the manuscript. LS thanks the ALMA/NAASC Staff for the support provided at the NA-ARC Headquarters at Charlottesville (US). LS acknowledges support from Fundaci\'on Marcos Moshinsky and UNAM PAPIIT Grant IN110122. LFM acknowledges support from grants PID2020-114461GB-I00, PID2023-146295NB-I00 and CEX2021-001131-S funded by MCIN/AEI/10.13039/501100011033, and by grant P20-00880, funded by the Economic Transformation, Industry, Knowledge and Universities Council of the Regional Government of Andalusia and the European Regional Development Fund from the European Union.  MAGM acknowledges the support of the State Research Agency (AEI) of the Spanish Ministry of Science and Innovation (MCIN) under grant PID2020-115758GB-I00/AEI/10.13039/501100011033. MAGM also acknowledges to be funded by the European Union (ERC, CET-3PO, 101042610). Views and opinions expressed are however those of the author(s) only and do not necessarily reflect those of the European Union or the European Research Council Executive Agency. Neither the European Union nor the granting authority can be held responsible for them. This article is based on work from the European Cooperation in Science and Technology (COST) Action NanoSpace, CA21126, supported by COST.

This paper makes use of the following ALMA data: ADS/JAO.ALMA\#2016.1.00944.S. ALMA is a partnership of the ESO (representing its member states), NSF (USA) and NINS (Japan), together with NRC (Canada), MOST and ASIAA (Taiwan), and KASI (Republic of Korea), in cooperation with the Republic of Chile. The Joint ALMA Observatory is operated by ESO, AUI/NRAO, and NAOJ. The National Radio Astronomy Observatory is a facility of the National Science Foundation operated under cooperative agreement by Associated Universities, Inc. This paper also makes use of the publicly available code {\sc DUSTY}.

\section*{DATA AVAILABILITY}

The data used in this article can be found in the ALMA public archives under the project 2016.1.00944.S.



\appendix

\end{document}